\documentclass[]{interact}

\usepackage[caption=false]{subfig} 
\usepackage[round]{natbib}
\newcommand{\mat}[1]{\mathbf{#1}} 

\usepackage{array}
\usepackage{lmodern} 
\usepackage[T1]{fontenc} 
\usepackage{amsmath} 
\usepackage{bm} 
\usepackage{float} 

\theoremstyle{plain} 

\theoremstyle{definition}

\theoremstyle{remark}

\usepackage{color}

\begin{document}

\articletype{ORIGINAL ARTICLE}

\title{Beta regression mixed model applied to sensory analysis}

\author{
\name{João César Reis Alves\textsuperscript{a}\thanks{CONTACT J.~C. ~R. Alves. Email: joaocesar@usp.br}; Gabriel Rodrigues Palma\textsuperscript{b}  and Idemauro Antonio Rodrigues de Lara\textsuperscript{a} }
\affil{\textsuperscript{a} University of São Paulo, "Luiz de Queiroz" College; \textsuperscript{b}Hamilton Institute, Maynooth University, Maynooth - Ireland}
}

\maketitle

\begin{abstract}
Sensory analysis is an important area that the food industry can use to innovate and improve its products. This study involves a sample of individuals who can be trained or not to assess a product using a hedonic scale or notes, where the experimental design is a balanced incomplete block design. In this context, integrating sensory analysis with effective statistical methods, which consider the nature of the response variables, is essential to answer the aim of the experimental study. Some techniques are available to analyse sensory data, such as response surface models or categorical models. This article proposes using beta regression as an alternative to the proportional odds model, addressing some convergence problems, especially regarding the number of parameters. Moreover, the beta distribution is flexible for heteroscedasticity and asymmetry data. To this end, we conducted simulation studies that showed agreement rates in product selection using both models. Also, we presented a motivational study that was developed to select prebiotic drinks based on cashew nuts added to grape juice. In this application, the beta regression mixed model results corroborated with the selected formulations using the proportional mixed model. 

\end{abstract}

\section{Introduction}
Sensory analysis is an essential scientific discipline for the accessibility or exclusion of foods, ranging from the evaluation of raw materials to the final result. The sensory analysis evaluates attributes such as colour, flavour, sweetness, overall impression, body, and aroma by utilising the senses of sight, touch, smell, and taste \citep{duan2024identification} and \citep{kottaridi2023regression}. These attributes are evaluated by people who may or may not be trained, and the resulting data is analysed statistically using different methodologies, such as response specifications or categorical models \citep{kumar2023impact}. In addition to providing an objective assessment of sensory characteristics, sensory analysis also contributes to the product's aesthetics, influencing demand and potential accessibility by the consumer public. However, considerations regarding panellists fatigue and associated costs become imperative when confronted with many treatments. In such contexts, employing incomplete blocks emerges as a viable strategy to streamline processes and ensure result accuracy \citep{shrikhande1965affine}.

One of the techniques used in the evaluation of sensory samples is the hedonic scales. Moreover, classification on categorical scales is widely used in sensory science and several other scientific disciplines where humans act as measuring instrumentscitep{christensen2013analysis}. The most common hedonic scale includes a score varying from 1 to 9, where “1” represents dislike extremely and “9” like extremely. The scores are given by panellists can be influenced by their effective memories and subjective assessments, which can introduce randomness and errors into the results \citep{sugumar2022comparative}. Thus, the experiment must be well planned and executed, using statistical methods that consider all possible errors and effects, respect the classification of variables, and ensure the robustness of the results.

In the literature, several studies treat the scores assigned on hedonic scales to sensory data as continuous variables to perform analysis of variance, assuming that the model is robust against violations of assumptions such as uncorrectable skew, heteroscedasticity and multimodality of the dependent variable. This view is often mistaken and can lead to erroneous interpretations \citep{smithson2006better}. For example, \citep{garcia2022sensory} aimed to provide the baking industry with a tool to design products tailored to consumer preferences; \citep{visalli2023hedonic} proposed a method to build a hedonic scale of attributes related to the perception of red wines; \citep{devi2023development} sought to improve the yield of reducing dietary fibre from pineapple pomace; \citep{sugumar2022comparative} performed a linear regression to establish the relationship between individual sensory configurations and the general acceptability of the samples, seeking to identify and understand the sensory configurations of Solanum nigrum leaf soup. \citep{garcia2022sensory} employed Principal Component Analysis to summarise descriptive analysis and identify differences in consumer acceptance of bread samples. On the other hand, \citep{melo2020modelos} explored mixed proportional odds models, which, though effective, encountered super parameterisation issues.  

In an increasingly competitive market, understanding consumer behaviour is crucial for the success of the food industry. Studying this behaviour allows for the development of products that meet consumer expectations, resulting not only in greater market success but also in reduced waste \citep{khan2023consumer}.
 
Hence, this work proposes using beta regression models with random effects to analyse sensory data. Beta regression models allow direct modelling of heteroscedastic and skewed data \citep{capelletti2024wind}. For scales with lower and upper limits, the beta distribution is a suitable candidate due to its flexibility \citep{kubinec2023ordered}. Moreover, these models eliminate prerequisites such as data normality and homogeneity of variance \citep{junaid2024modified}, which reduces super parameterisation problems.  
 
A simulation study was carried out to illustrate the efficiency of this model, initially adopting cumulative logit models with proportional odds configurations and, subsequently, testing the potential of the beta regression model. It is expected, therefore, that the beta regression models with a random effect of the panellists will present good results, contributing as a tool for research in sensory analysis and helping in the selection of increasingly functional and innovative products for the consumer.

\section{Review of beta regression Model with Random Effect} \label{rev2}

According to the authors \citep{figueroa2013mixed} and \citep{zerbeto2014melhor}, beta regression models with random effects are extension cases of Generalised Linear Mixed Models (GLMM). In GLMM, the linear predictor is formed by fixed effects and random effects, which are incorporated into the model to consider the correlated observations \citep{noguera2019approximate}, resulting from repeated measurements for individuals \citep{clayton1996generalized}. 
In this context, let a longitudinal study, in which $\mat{y}_{i}=(y_{i1}, \ldots, y_{in_{i}})^{T}$ is the vector with $n_{i}$ observations, $y_{ij}$, of the \textit{i}-th individual, $i= 1, 2, \ldots, N$. Also, let $\mat{x}_{ij}=(x_{ij1}, \ldots, x_{ijp} )^{T}$ the  matrix $(n_{i}\times p)$, corresponding to the vector $\bm{\beta}=(\beta_{1}, \ldots, \beta_{p})^{T}$ of unknown regression coefficients related to fixed effects;
$\mat{z}_{ij} = (z_{ij1}, \ldots, z_{ijq} )^{T}$ is the matrix  $(n_{i}\times q)$ associated to the vector $\mat{u}_{i}=(u_{i1}, \ldots, u_{iq})^{T}$ related to random effects. Then, supposing that $Y_{ij}|\mat{u}_{i}$ are conditionally independent random variables with beta distribution$~(\mu_{ij},\phi)$;
 $u_{1},...,u_{q}$ independent random variables with distribution $\mat{u}_{i}|\mat{D} \sim N_{q}(\mat{0},\mat{D})$, where $\mat{D} $ is a positive defined matrix \citep{bonatmodelo}, 
the mixed beta regression model can be defined as:
\begin{equation} g(\mbox{E}(\mat{Y}_{i}|\mat{u}_{i}))=\mat{x}_{i}^{T}\bm{\beta} +\mat{ z}_{i}^{T}\mat{u}_{i}=\eta_{i} \label{betamista}
\end{equation}
where $g(.)$ is a known, strictly monotone, doubly differentiable link function. It is usual but not restricted to use  the logit link function (\cite{zimprich2010modeling}), then we rewrite the equation (\ref{betamista}) as: 
\begin{equation}
      \ln\left ( \frac{\mu _{ij}}{1-\mu _{ij}} \right )=\eta _{ij}=\mat{x}_{ij}^{T}\bm{\beta} +\mat{z}_{ij}^{T}\mat{u}_{i}, \label{liga}
\end{equation}\\
and $\mu _{ij}=\mbox{E}(Y_{ij}|\mat{u}_{i})$ by means equation (\ref{liga}) is given by:\\
\begin{equation*}
    \mu _{ij}=\frac{\exp(\eta _{ij})}{1+\exp(\eta _{ij})}
\end{equation*}

To estimate the parameters of the mixed beta regression  (equation \ref{betamista}), the iterative likelihood method was used. The contribution to the likelihood function from individual $i$ is given by
\begin{equation}
    f_{i}(\mathbf{y}_{i}|\bm{\beta} , \mathbf{D}, \phi )=\int \prod_{j=1}^{n_{i}}f_{ij}(y_{ij}|\mathbf{u}_{i},\bm{\beta} , \phi )f(\mathbf{u}_{i}|\mathbf{D})\textit{d}\mathbf{u}_{i}, 
    \end{equation}
where the likelihood for $\bm{\beta}, \mathbf{D}$ and $\phi$, is 

\begin{equation}
    L(\bm{\beta} , \mathbf{D}, \phi )=\prod_{i=1}^{N}f_{i}(\mathbf{y}_{i}|\bm{\beta} , \mathbf{D}, \phi )
=\prod_{i=1}^{N}\int \prod_{j=1}^{n_{i}}f_{ij}(y_{ij}|\mathbf{u}_{i},\bm{\beta} ,\phi )f(\mathbf{u}_{i}|\mathbf{D})\textit{d}\mathbf{u}_{i}. \label{41}
\end{equation}

The main problem in maximising the equation (\ref{41}) is the presence of $N$ integrals over the $q$-dimensional random effects $\mathbf{u}_{i}$, which does not have a closed form of solution and, in general, constitutes the key to the estimation process \citep{bates2009package}. Therefore, the Laplace approximation is considered as the standard method for solving these integrals \citep{brooks2017glmmtmb}. This method is implemented in the package \texttt{glmmTMB} \citep{magnusson2017package}, available in the software \texttt{R} \citep{r2013r}.  More details on the model estimates using the\texttt{TMB} package, see \citep{kristensen2015tmb} and  other estimation methods can be found in \citep{verkuilen2012mixed}, \citep{manco2013modelos}, \citep{noguera2019approximate}, \citep{zerbeto2014melhor}, \citep{molenberghs2005joint}.

\section{Methods}
\label{methods}

Considering a sensory study according to a balanced incomplete block design with $2^{2}$ factorial central rotational composite \citep{myers2016response}, i.e., in which each panellists evaluates $k$ products of $v$ available, each of them being repeated $r$ times, where a pair of varieties occurs a $\lambda$ number of times in the same block. In a balanced incomplete block design, the following conditions must be met \citep{arruda1955delineamento}:

\begin{equation*}
    bk=rv, \label{equação1}
\end{equation*}
Number of variety pairs
\begin{equation*}
    	  \frac{(v(v-1))}{2}, \label{equação2}
\end{equation*}
Number of blocks
\begin{equation*}
    		 \frac{(k(k-1))}{2}, \label{equação3}
\end{equation*}
Number across the entire experience
\begin{equation*}
    		 \frac{bk(k-1)}{2}=\frac{\lambda v(v-1)}{2}, \label{equação4}
\end{equation*}

In the last equality, replacing $bk$ with $rv$, we have:
\begin{equation}
\lambda(v-1)= r(k-1), \label{equação5}
\end{equation}
To apply the process described in the section \ref{rev2}, initially the  $y_{i}$, was converted to the $(0,1)$. For this, we have used the equation proposed by \citep{smithson2006better}:
\begin{equation}\label{converte}
y^*_{i}=\frac{(y_{i}(n-1)+0.5)}{n},
\end{equation}
where $n$ is the observations sample size. After that, the data can be considered as sample realisations of continuous variable with restriction borders \citep{sarzo2023modelling}, according to the class of beta regression models as proposed by \citep{ferrari2004beta}.

To evaluate the effect of the product on a sensory attribute, we considerer three models. The null is:
\begin{equation}\label{nulo}
     \eta_{0}=\log\left (\frac{\mu _{i}}{1-\mu _{i}}  \right ) =\alpha_{0}, 
\end{equation}
representing the model in which the effect of any sensory attribute is independent of the formulation. The next one includes the formulation~(product) effect:
\begin{equation}\label{produto}
     \eta_{1}=\log\left (\frac{\mu _{i}}{1-\mu _{i}}  \right ) =\alpha_{0}+\mathbf{x}_{i}^{T}\bm{\beta},
\end{equation}
on what $\mathbf{x}_{i}^{T}$ represents the design matrix  and $\bm{\beta} = (\beta_{1}, \ldots, \beta_{p})^{T}$ the associated vector, that represents the product effect on the attribute response, when we compare with the model \ref{nulo}. Finally, including random effects because the structure of incomplete block and uncontrollable effect panellists:  
\begin{equation}
\label{misto}
    \eta_{2}=\log\left (\frac{\mu _{i}}{1-\mu _{i}}  \right ) =\alpha_{0}+\mathbf{x}_{i}^{T}\bm{\beta}+\mathbf{u}_{i},
\end{equation} 
assuming $\mathbf{u}_{i} \sim N(\mathbf{0},\sigma^{2})$. For model selection, we used the Maximised Likelihood procedure and the Akaike Information Criterion. Regardless of the linear predictor, $i=0,1,2$ the estimated mean values are given by the equation:
\begin{equation}\label{equation10}
    \hat{\mu_{i}}= \frac{\mbox{exp}(\eta_{i})}{1-\mbox{exp}(\eta_{i})}
\end{equation}

\section{Simulation studies}
\label{simula}

A simulation study was conducted to evaluate the beta regression model's performance in the sensory data analysis. This process was carried out with three formulations, using balanced complete blocks to enable the simulation. Our focus was not to explore different designs but to evaluate the model's effectiveness compared to the commonly used cumulative logit models with proportional odds. To simulate the data, parameters from the provided mixed cumulative logit model with proportional odds, represented by:
\begin{equation}\label{modelologito}
    \eta _{i}= ln \left [ \frac{\gamma  _{j}}{1-\gamma _{j}} \right ]=\alpha _{j}+\bm{\beta}^{T}\mathbf{X}+\mathbf{u}_{i},
\end{equation}
with $\alpha_{j}$ being the intercept of the $j-th$ response category referring to a given sensory attribute (global impression, colour, aroma, sweetness, flavour and body), $\bm{\beta}$ the vector of regression parameters associated with the $\mathbf{X}$ design matrix for fixed effects, and $\mathbf{u}_{i}$ is the random effect associated with the $i-th$ panellists, where, $\mathbf{u}_{i}\sim N(0,\sigma _{u}^{2})$.  The parameters value used in the simulations process are presented in the Table \ref{parametroslogitos}. The reference category for the response was "1 = dislike extremely" and the formulation was $F_{1}$.

\begin{table}[H]
	\centering
\caption{Fixed parameters of the proportional probability model, used to simulate data in sensory analysis considering thirteen scenarios.}
\label{parametroslogitos}
\label{SN}
\begin{tabular}{clcl}

\\ \hline
Scenarios & Parameters (logits) & Scenarios & Parameters(logits)\\
          & $\bm{\theta}_{j}=(\alpha_{j}, \beta_{[f2]},\beta_{[f3]})$ &  & $\bm{\theta}_{j}=(\alpha_{j}, \beta_{[f1]},\beta_{[f2]})$
\\\hline
$F_{1}=F_{2}=F_{3}$ & $\bm{\theta}_{1}=(5, 5, 5)$ &
$F_{3}=F_{1}<F_{2}$   &
$\bm{\theta}_{1}=(5, 5, 0)$ \\
&
$\bm{\theta}_{2}=(0, 0, 0)$ &
&
$\bm{\theta}_{2}=(0, 0, 0)$ \\
 &
$\bm{\theta}_{3}=(0, 0, 0)$&
&
$\bm{\theta}_{3}=(0, 0, 6)$ \\
&
$\bm{\theta}_{4}=(0, 0, 0)$ &
&
$\bm{\theta}_{4}=(0, 0, 0)$ \\ \hline
  
$F_{1}=F_{2}<F_{3}$ &
  $\bm{\theta}_{1}=(5, 5, 0)$ &
  $F_{3}<F_{1}=F_{2}$ &
  $\bm{\theta}_{1}=(5, 0, 0)$ \\
 &
$\bm{\theta}_{2}=(0, 0, 0)$ &
   &
$\bm{\theta}_{2}=(0, 0, 0)$ \\
 &
$\bm{\theta}_{3}=(0, 0, 0)$ &
   &
$\bm{\theta}_{3}=(0, 6, 6)$ \\
 &
$\bm{\theta}_{4}=(0, 0, 7)$& 
 & 
 $\bm{\theta}_{4}=(0, 0, 0)$\\
 \hline 
$F_{1}<F_{2}=F_{3}$ &
  $\bm{\theta}_{1}=(5, 0, 0)$ &
  $F_{3}<F_{1}<F_{2}$ &
  $\bm{\theta}_{1}=(5, 0, 0)$ \\
 &
  $\bm{\theta}_{2}=(0, 0, 0)$ &
   &
  $\bm{\theta}_{2}=(0, 0, 0)$ \\
 &
 $ \bm{\theta}_{3}=(0, 6, 6)$ &
   &
  $\bm{\theta}_{3}=(0, 6, 0)$ \\
 &
  $\bm{\theta}_{4}=(0, 0, 0)$ &
   &
  $\bm{\theta}_{4}=(0, 0, 7)$ \\ 
  \hline 
  & Parameters(logits) &  & Parameters(logits)\\
          & $\bm{\theta}_{j}=(\alpha_{j}, \beta_{[f2]},\beta_{[f3]})$&  & $\bm{\theta}_{j}=(\alpha_{j}, \beta_{[f3]},\beta_{[f2]})$\\
          
          \hline
 $F_{1}<F_{2}<F_{3}$ & $\bm{\theta}_{1}=(5, 0, 0)$ &
$F_{1}<F_{3}<F_{2}$   &
$\bm{\theta}_{1}=(5, 0, 0)$ \\
&
$\bm{\theta}_{2}=(0, 0, 0)$ &
&
$\bm{\theta}_{2}=(0, 0, 0)$ \\
 &
$\bm{\theta}_{3}=(0, 6, 0)$&
&
$\bm{\theta}_{3}=(0, 6, 0)$ \\
&
$\bm{\theta}_{4}=(0, 0, 7)$ &
&
$\bm{\theta}_{4}=(0, 0, 7)$ \\ \hline 
 & Parameters (logits) &  & Parameters(logits)\\
          & $\bm{\theta}_{j}=(\alpha_{j}, \beta_{[f1]},\beta_{[f3]})$ &  & $\bm{\theta}_{j}=(\alpha_{j}, \beta_{[f3]},\beta_{[f1]})$\\
          
          \hline
  $F_{2}<F_{1}=F_{3}$ & $\bm{\theta}_{1}=(5, 0, 0)$ &
$F_{2}=F_{3}<F_{1}$   &
$\bm{\theta}_{1}=(5, 5, 0)$ \\
&
$\bm{\theta}_{2}=(0, 0, 0)$ &
&
$\bm{\theta}_{2}=(0, 0, 0)$ \\
 &
$\bm{\theta}_{3}=(0, 6, 0)$&
&
$\bm{\theta}_{3}=(0, 0, 6)$ \\
&
$\bm{\theta}_{4}=(0, 0, 0)$ &
&
$\bm{\theta}_{4}=(0, 0, 0)$ \\ \hline

  $F_{2}<F_{1}<F_{3}$ & $\bm{\theta}_{1}=(5, 0, 0)$ &
$F_{2}<F_{3}<F_{1}$   &
$\bm{\theta}_{1}=(5, 0, 0)$ \\
&
$\bm{\theta}_{2}=(0, 0, 0)$ &
&
$\bm{\theta}_{2}=(0, 0, 0)$ \\
 &
$\bm{\theta}_{3}=(0, 6, 0)$&
&
$\bm{\theta}_{3}=(0, 6, 0)$ \\
&
$\bm{\theta}_{4}=(0, 0, 7)$ &
&
$\bm{\theta}_{4}=(0, 6, 0)$ \\ 
   \hline
   & Parameters (logits) &  & \\
          & $\bm{\theta}_{j}=(\alpha_{j}, \beta_{[f2]},\beta_{[f1]})$ &
          \\ \hline
$F_{3}<F_{2}<F_{1}$ &
 
  $\bm{\theta}_{1}=(5, 0, 0)$ 
   &
  \\
 &
  $\bm{\theta}_{2}=(0, 0, 0)$ &
  &
  \\
 &
  $\bm{\theta}_{3}=(0, 6, 0)$ &
  \\
 & $\bm{\theta}_{4}=(0, 0, 7)$ 
 \\ \hline

\end{tabular} 
\end{table}

Once the data were generated, the cumulative logit models with proportional odds and the beta regression model with random effects were fitted. For the models proportional odds (equation \ref{modelologito}), the function \texttt{clmm2(.)} from the \texttt{ordinal} \citep{christensen2015package} package was used. In the beta regression model with random effects (equation \ref{misto}), as described in section \ref{methods}. Next, the method was implemented using the package \texttt{glmmTMB} \citep{magnusson2017package, brooks2017glmmtmb}. For each scenario described in Table \ref{parametroslogitos}, $1,000$ data simulations were performed, each with $N=90$ and $N=300$ panellists (blocks). The results of the agreement rates between the two fitted models are presented in Table \ref{tabela:simulation}. 

\begin{table}[H]
\begin{center}
\caption{Agreement rates between cumulative logit model with proportional odds and the beta regression model with random effects for the thirteen different simulation study scenarios, considering the three formulations ($F_{1}$, $F_{2}$ and $F_{3}$ ), for $N=90$ and $N=300$.}
\label{tabela:simulation}

\begin{tabular}{lrr>{\raggedleft\arraybackslash}p{3cm}}
\hline
\\ Scenarios    & \multicolumn{2}{c}{Agreement rates} \\ 
              &   N=90 & N=300     \\ \midrule 
$F_{1}=F_{2}=F_{3}$  & 98.2\% & 87.1\% \\
$F_{1}=F_{2}<F_{3}$  & 99.1\% & 99.4\% \\
$F_{1}<F_{2}=F_{3}$  & 91.0\% & 94.4\% \\
$F_{1}<F_{2}<F_{3}$  & 99.9\% & 100.0\% \\
$F_{2}<F_{1}=F_{3}$  & 91.6\% & 92.6\% \\
$F_{2}<F_{1}<F_{3}$ & 99.9\% & 100.0\% \\
$F_{3}=F_{1}<F_{2}$  & 100.0\% & 98.8\% \\
$F_{3}<F_{1}=F_{2}$ & 92.3\% & 93.3\% \\
$F_{3}<F_{1}<F_{2}$ & 99.9\% & 100.0\% \\
$F_{1}<F_{3}<F_{2}$  & 99.9\% & 100.0\% \\
$F_{2}=F_{3}<F_{1}$  & 99.7\% & 98.2\% \\
$F_{2}<F_{3}<F_{1}$  & 99.7\% & 100.0\% \\
$F_{3}<F_{2}<F_{1}$  & 99.9\% & 100.0\% \\ \hline
\end{tabular}
\end{center}
\end{table}

The agreement rates between the cumulative logit models with proportional odds and the proposed method are presented in the Table \ref{tabela:simulation}. The results show that for $N = 90$ the agreement rates between the models \ref{modelologito} and \ref{misto}, were equal to or greater than $91.0\%$ in all scenarios. Notably, there is a satisfactory agreement rate between the models, with this rate increasing in all scenarios from $N = 90$ to $N = 300$, except in scenarios where $F_{2}=F_{3}<F_{ 1}$, $F_{1}=F_{2}=F_{3}$ and $F_{3}=F_{1}<F_{2}$. This phenomenon can be attributed to randomness and the presence of random effects.

Data analysis reveals that the performance of specific scenarios, such as $F_{1}<F_{3}<F_{2}$, achieved an agreement rate of $99.9\%$ for $N=90$ and $100\%$ for $N = 300$. This high performance was repeated in other scenarios evaluated, demonstrating that, in general, the results were promising. Based on the simulation study results, it can be concluded that the  beta regression model with random effects model demonstrated a high efficiency in agreement rate when analysing sensory data.

\section{Applications and Results}

As an application, we have used a dataset according to sensory study, that was developed at the Department of Food Technology at the Federal University of Ceará (UFC), in 2016 \citep{rebouccas2016bebida}, followed a balanced incomplete block design, in which each of the $130$ untrained panellists evaluated $4$ of the $13$ proposed beverage formulations. The drinks were made with prebiotic beverages from cashew nut added to grape juice, following a central composite design. 
This design consists of four points at the corners of a square, five points in the centre of this square and four axial points. In terms of the encoded variables, the corners of the square are (\% sugar, \% juice) = $(4, 20), (4, 40), (8, 20), (8, 40)$. The centre points are at (\% sugar, \% juice) = $(6, 30)$, and the axial points are at (\% sugar, \% juice) = $(3, 30), (9, 30 ), (6, 16) , (6, 44)$. 

Originally, the panellists evaluated each of its 4 formulations using a hedonic scale that is composed by $9$ ordinal points, that is, from the least favourable note to the one with the greatest representative value. The sensory attributes that were evaluated: overall impression, aroma, colour, body, sweetness and flavour. Analyses using the original assessment scale, in general, can be done via the proportional odds model \citep{agresti2010modeling}, which, in general, can present convergence problems due to an excess of parameters, as in studies developed by \citep{melo2020modelos}, which needed to reduce the scale to $5$ points, to obtain convergence of the model. Alternatively, in this work, the evaluation scale is adjusted to the interval $(0,1)$, as described by the equation \ref{converte}, allowing analysis through the beta regression model. Additionally, in the new data scale $(0, 1)$, we present a brief exploratory analysis to visualize the response pattern for each of the evaluated attributes.

\begin{figure}[!htbp]
\centering \hspace{-2em}
  \includegraphics[scale=0.38]{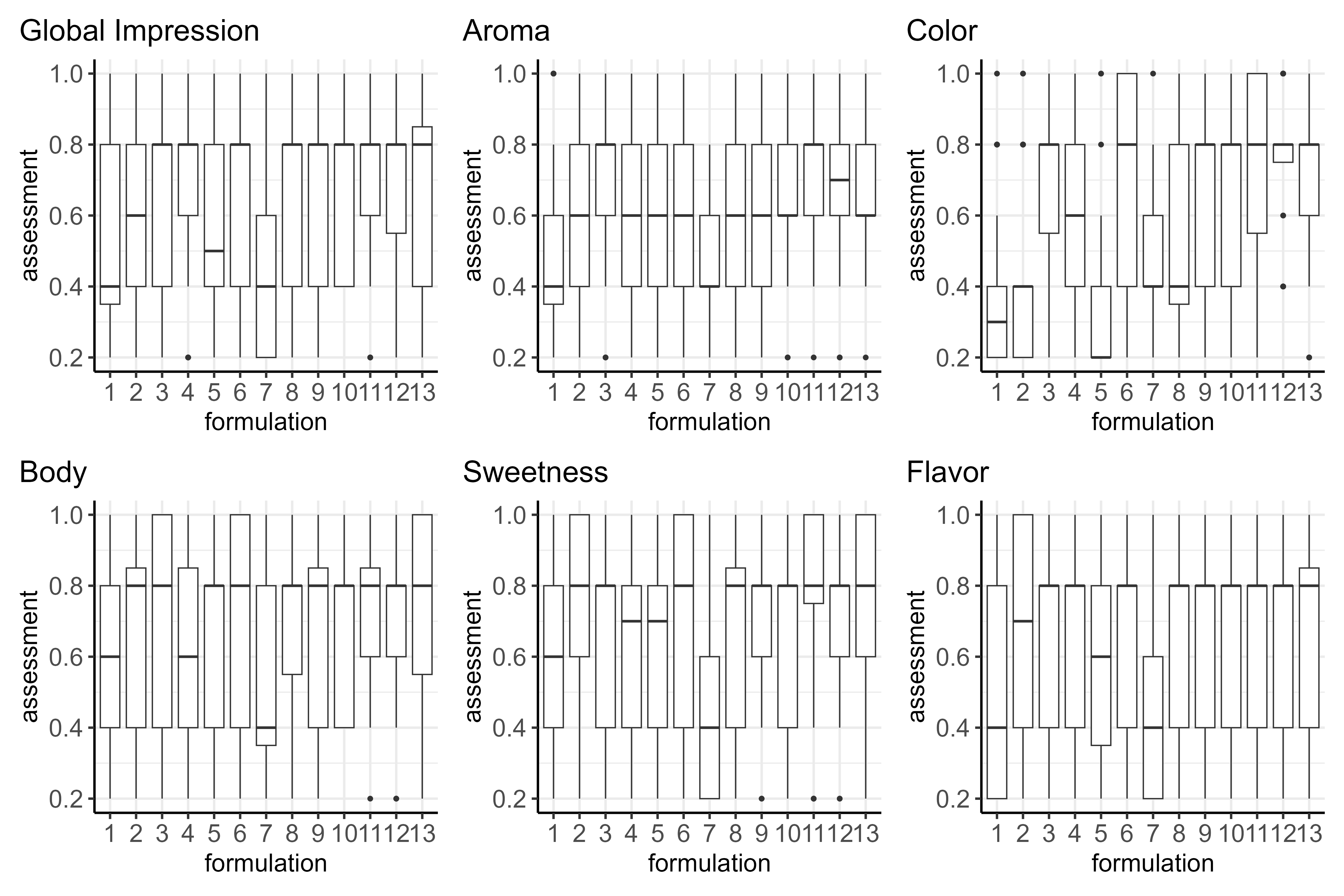}
  \caption{Boxplots of the sensory attributes for the 13 formulations of prebiotic grape juice drinks, according to conducted at UFC in 2016.} 
  \label{boxplot}
\end{figure}

As illustrated in Figure \ref{boxplot}, data variability and asymmetry are observed for all sensory attributes, except in formulation $F_{2}$ for the overall impression and aroma attributes. Outliers were identified in all sensory attributes except for flavour. For example, in the colour attribute, outliers were observed in formulations $F_{1}$, $F_{2}$, $F_{5}$, $F_{7}$, $F_{12}$, and $F_{13}$. These outliers can indicate the presence of additional variability and possible external influences in the data analysis.The formulations $F_{6}$, $F_{13}$ and $F_{4}$ stand out for presenting higher median values in several sensory attributes, such as overall impression, colour, body, sweetness and flavour. In contrast, the formulations $F_{1}$, $F_{7}$ and $F_{5}$ exhibit the lowest median values.

\newpage

However, this analysis is preliminary and is only exploratory in nature. Next, as described in the methodology,  beta regression models were fitted to data for all sensory attributes. Null model (equation \ref{nulo}) considers only the intercept, Model 1 (equation \ref{produto}) that includes the formulation effect and, finally, model 2~(equation \ref{misto}) that is complete, that is, formulation fixed effect and random effect for the panellist. Also, to select the best structure, we considered the estimated dispersion parameter ($\hat{\phi}$), the logarithm of the maximised likelihood function ($\log(L)$) and Akaike information criterion (AIC), that are presented in the Table~\ref{tabela:2}.

\begin{table}[h]
\centering
\caption{Estimated dispersion parameter ($\hat{\phi}$), the logarithm of maximized likelihood function ($\log(L)$) and Akaike information criterion (AIC) for all sensory attributes, considering the three models: Null (equation~\ref{nulo}), Formulation effect (equation~\ref{produto}) and with Formulation and random effect (equation~\ref{misto}) referring to the study carried out at UFC in 2016.}
    \label{tabela:2}
 \vspace{0.2cm}   
\begin{tabular}{l lccc}
\hline
{Sensory Attribute} &     Models     & $\hat{\phi}$   &  $\log(L)$   & AIC    \\ \hline
{Global impression}       & Null~(intercept) & 1.8 & 128.1 & -252.1 \\
                                        &$1$~(intercet + formulation effect) & 2.0 & 148.4 & -268.7 \\
                                        & $2$~(intercet + formulation+random effects) & 4.4 & 232.5 & -435.0 \\ \hline
{Aroma}                  & Null~(intercept) & 2.1 & 81.7  & -158.7 \\
                                        & $1$~(intercet + formulation effect) & 2.3 & 96.6  & -165.3 \\
                                        & $2$~(intercet + formulation+random effects) & 6.3 & 229.5 & -429.0 \\ \hline
{Body}                  & Null~(intercept) & 1.6 & 237.2 & -470.4 \\
                                        & $1$~(intercet + formulation effect) & 1.7 & 250.8 & -473.7 \\
                                        & $2$~(intercet + formulation+random effects) & 4.1 & 343.9 & -657.8 \\ \hline
{Sweetness}                 & Null~(intercept)& 1.6 & 254.4 & -504.8 \\
                                        & $1$~(intercet + formulation effect) & 1.7 & 271.8 & -515.6 \\
                                        & $2$~(intercet + formulation+random effects) & 3.1 & 327.2 & -624.3 \\ \hline
{Flavour}                  & Null~(intercept) & 1.6 & 139.5 & -275.0 \\
                                        & $1$~(intercet + formulation effect) & 1.7 & 153.7 & -279.4 \\
                                        & $2$~(intercet + formulation+random effects) & 3.6 & 237.0 & -444.0 \\ \hline
{Colour}                    & Null~(intercept) & 1.5 & 121.5 & -239.0 \\
                                        & $1$~(intercet + formulation effect) & 1.9 & 179.1 & -330.1 \\
                                        & $2$~(intercet + formulation+random effects) & 4.8 & 286.8 & -543.5 \\ \hline

\end{tabular}
\end{table}

Based on the indicators presented in Table \ref{tabela:2}, it is observed that both the logarithm of the likelihood function and the estimate of the dispersion parameter increase towards the complete model, indicating that the variability is better explained by the mixed model (equation \ref{misto}) for all sensory attributes. Furthermore, including the random effect for the taster is pertinent in terms of the incomplete block design and corroborates the observed AIC values (the lowest are for the mixed model in all attributes). The likelihood ratio test~(p-value $<0.01$) for all attributes) confirm that the model given by the equation~\ref{misto} best describes the function structure of the data. Also, the formulation effect was significant for all attributes  (p-value $<0.05$), considering a significance level of 95\%. In this context, the estimated values for the variance of the random effect of panellist were: $\hat{\sigma }_{GI}^{2}=0.97$, $\hat{\sigma }_{aroma}^{2}=1.00$, $\hat{\sigma }_{body}^{2}=1.08$, $\hat{\sigma }_{sweetness}^{2}=0.89$, $\hat{\sigma }_{flavour}^{2}=1.00$, $\hat{\sigma }_{colour}^{2}=1.05$.

\begin{table}[H]
\centering
\caption{Means estimated values by the equation \ref{equation10} according to the sensory attributes and formulations, resulting from the study carried out at UFC in 2016.}
\label{tabela:3}

\vspace{0.2cm }
\begin{tabular}{lrrrrrr}
\cline{1-7}
                             &Global Impression & Aroma& Body & Sweetness & Flavour & Colour\\ \cline{1-7}
Formulation 1  & 0.5014   &     0.5429   &  0.6028  & 0.6386 &  0.5442   & 0,3816           \\
Formulation 2  & 0.6357    &    0.6306   &  0.7145 &  0.7570 &   0.6664&  0.3830     \\
Formulation 3  & 0.6993   &     0.6781  &  0.7153 &    0.6877 & 0.6742&0.7235\\
Formulation 4  & 0.7817    &    0.6770   & 0.7475&    0.7454  &  0.7054 & 0.7204\\
Formulation 5  & 0.5837    &    0.6108  & 0.7127&  0.7014 &     0.6510 &  0.3134\\
Formulation 6 & 0.7833     &  0.7621   & 0.8040 &  0.8136 & 0.7514& 0.7998   \\
Formulation 7  & 0.5647    & 0.5700   &  0.6222 &   0.5670 & 0.5353 & 0.5455\\
Formulation 8  & 0.7140    &  0.6487  &  0.7624 &   0.7744  & 0.7081& 0.5948\\
Formulation 9  & 0.7154    &   0.6567 &  0.8009 &   0.7311 & 0.6977& 0.7752\\
Formulation 10 & 0.6827    &   0.6717 &  0.7226&  0.7144  &   0.6683& 0.7454       \\
Formulation 11 & 0.7139    & 0.7067  &   0.7393 &   0.8045  &0.7032&0.7915\\
Formulation 12 & 0.7299    &  0.6879 &  0.7500 &    0.7300 & 0.6922& 0.7977\\
Formulation 13 & 0.7922    &  0.7135 &  0.8153 &    0.7998 & 0.7614&0.7478\\ \cline{1-7}
\end{tabular}
\end{table}

The Table \ref{tabela:3}, it can be observed that the prebiotic formulations of the central point (F9 to F13) present estimated mean  values varying  from 0.6567 to 0.8153 for all sensory attributes. Among these, formulation F13, central point, (6\% sugar and 30\% grape juice) stands out, with estimated mean of 0.7922 for global impression, 0.7135 for aroma, 0.8153 for body, 0.7998 for sweetness, 0.7614 for flavour and 0.7478 for colour.

It is also observed that formulations F4 (8\% sugar and 40\% grape juice) and F6 (6\% sugar and 44\% grape juice) presented the highest estimated values, respectively: 0.7817 and 0.7833 for overall impression; 0.6770 and 0.7621 for aroma; 0.7475 and 0.8040 for body; 0.7454 and 0.8136 for sweetness; 0.7054 and 0.7514 for flavour; and 0.7204 and 0.7998 for colour. Formulation F6 (6\% sugar and 44\% grape juice) was the best accepted in relation to all sensory attributes, while formulation F1 (3\% sugar and 30\% grape juice) was the least accepted. These conclusions are consistent with the observations of \citep{melo2020modelos}, who used mixed proportional likelihood models.

\section{Conclusion}

This study presents beta regression models with random effects for data analysis based on sensory studies, in which the response variable can be considered on a scale of $(0, 1)$. Compared to the logit model, the advantage of the proposed model is its greater flexibility, reducing problems with superparameterisation. The results highlighted the importance of formulation covariates and the random effect of the panellists for a more accurate model. The findings derived from our simulation study consistently indicate that, in the scenarios examined, the proposed model achieved a high agreement rate with logit proportional odds models, constituting an alternative analysis for researchers and professionals in the food industry.

Although in the application considered in this study, all data is contained in the open interval $(0, 1)$, it is essential to highlight that it is possible to model variables within the closed interval $[0, 1]$. For future research, carrying out simulations using Balanced Incomplete Block (BIB) designs would also be relevant. Furthermore, it is essential to investigate other link functions besides the logit function. Conducting additional analyses of the beta regression model's performance, particularly with random effects, is a critical step in our research. This includes a thorough evaluation of residuals and diagnoses.

\section{Acknowledgments}
This publication has emanated from research conducted with the financial support of Brazilian Fundation, Coordenação de Aperfeiçoamento de Pessoal de Nível Superior (CAPES)  process number (88887.821274/2023-00) and Science Foundation Ireland under Grant number 18/CRT/6049.

\bibliography{referencias.bib} 

\end{document}